\documentclass[a4paper,twoside]{article}
\newcommand{\affil}[1]{$^{\rm #1}$}
\usepackage[authoryear]{natbib}
\bibpunct{(}{)}{;}{a}{}{,}
\usepackage{graphicx}
\date{} 
\newcommand{\kms}{\mbox{km\,s$^{-1}$}}
\newcommand{\water}{\mbox{H$_2$O}}
\newcommand{\hii}{\mbox{H{\scriptsize II}}}
\newcommand{\arcsec}{\mbox{$^{\prime\prime}$}}
\title{\large\bf\flushleft Maser Source Finding Methods in HOPS}
\author{\parbox{\textwidth}{\flushleft
\vspace{-0.5cm}
{\it Andrew J. Walsh\affil{A,F}, Cormac Purcell\affil{B}, Steven Longmore\affil{C}, Christopher H. Jordan\affil{A,D}, and Vicki Lowe\affil{D,E}}\\
\vspace{0.4cm}
{\small \affil{A}\,Centre for Astronomy, School of Engineering and Physical Sciences, James Cook University, Townsville, QLD, 4814.}\\
{\small \affil{B}\,School of Physics and Astronomy, University of Leeds, Leeds, LS2 9JT, UK}\\
{\small \affil{C}\,European Southern Observatory, Karl-Schwarzschild-Str. 2, 85748 Garching, Germany}\\
{\small \affil{D}\,CSIRO Astronomy and Space Science, PO BOX 76, Epping, NSW, 1710}\\
{\small \affil{E}\,School of Physics, University of NSW, Sydney, NSW, 2052}\\
{\small \affil{F}\,Email: andrew.walsh@jcu.edu.au}}}
\begin{document}
\twocolumn[
\begin{changemargin}{.8cm}{.5cm}
\begin{minipage}{.9\textwidth}
\vspace{-1cm}
\maketitle
%
%
\small{\bf Abstract:}
The {\bf H}$_2${\bf O} Southern Galactic {\bf P}lane {\bf S}urvey (HOPS) has
observed 100 square degrees of the Galactic plane, using the Mopra radio telescope
to search for emission from multiple spectral lines in the 12\,mm band (19.5\,--\,27.5\,GHz).
Perhaps the most
important of these spectral lines is the 22.2\,GHz water maser transition. We describe the
methods used to identify water maser candidates and subsequent confirmation of
the sources. Our methods involve a simple determination of likely candidates by
searching peak emission maps, utilising the intrinsic nature of water maser
emission - spatially unresolved and spectrally narrow-lined. We estimate
completeness limits and compare our method with results from the {\sc Duchamp} source finder.
We find that the two methods perform similarly. We conclude that the similarity in
performance is due to the intrinsic limitation of the noise characteristics of the data.
The advantages of our method
are that it is slightly more efficient in eliminating spurious detections and is simple
to implement. The disadvantage is that it is a manual method of finding sources and
so is not practical on datasets much larger than HOPS, or for datasets with extended emission
that needs to be characterised. We outline a two-stage method
for the most efficient means of finding masers, using {\sc Duchamp}.


\medskip
\medskip
\end{minipage}
\end{changemargin}
]
\small

\section{Introduction}
Astrophysical masers have been known for over forty years
\citep{weaver65,gundermann65}. They are known to occur in transitions of many
chemical species, with the most common seen in hydroxyl (OH), methanol
(CH$_3$OH), silicon monoxide (SiO) and water (H$_2$O). The 22.2\,GHz
\water~maser transition
is the strongest and most widespread known to date. It is seen towards a
variety of astrophysical objects such as low- and high-mass star forming regions
(eg. \citealt{cheung69,forster99,claussen96}) evolved stars
\citep{dickinson76,miranda01,hinkle79,barlow96} and the centres
of active galaxies \citep{claussen84}. Other masers are often seen arising from
the same sites that give rise to \water~masers. In particular, OH and CH$_3$OH
masers occur within the spatial resolution of observations (typically an arcsecond
or less) of \water~masers in high-mass star forming regions
\citep{forster00,walsh98}. This has led to the identification of a tentative
evolutionary sequence where both CH$_3$OH and \water~masers appear early on during
the formation process, with OH masers appearing at a more evolved stage, but with
some overlap between all three maser species. Such arguments are based
on the statistical occurrence of the masers towards other signposts of high-mass
star formation, such as ultracompact \hii~regions and hot cores. However, to
date most searches for these masers have targeted known regions of star formation
and thus may suffer from biases. In order to reduce such biases of candidate
selection, it is important to conduct untargeted surveys of the Galaxy for these
masers.

The methanol multibeam project \citep{caswell10} is one such survey, which
focuses on the strongest and most widespread of the Class II CH$_3$OH
masers\footnote{Class I CH$_3$OH masers are collisionally pumped and are usually associated
with weak shocks whereas Class II CH$_3$OH masers are radiatively pumped and closely
associated with the early stages of high mass star formation (eg. \citealt{voronkov05}
and references therein).} at
6.7\,GHz. This survey is designed to cover the entire Galactic plane in a 4$^\circ$
wide latitude band. In the near future, there will be a survey on the Australian SKA
Pathfinder radiot elescope (ASKAP), called GASKAP, which will survey the southern
Galaxy up to 10$^\circ$ away from the Galactic plane. One of the target spectral
species for GASKAP is OH, including the main maser lines at 1.665 and 1.667\,GHz.
The \water~maser line has recently been surveyed as part of the
{\bf H}$_2${\bf O} southern Galactic {\bf P}lane {\bf S}urvey (HOPS), using the Mopra
radiot elescope \citep{walsh11}. HOPS has observed 100 square degrees of the southern
Galactic plane at 12\,mm and includes emission from the 22.2\,GHz \water~maser as
one of its main target lines. HOPS forms the basis for the work reported in this paper. 

With the advent of these untargeted, large scale surveys, the problem of
finding and identifying new sources has become an important challenge. HOPS
is the smallest of the surveys mentioned above, in terms of equivalent beam
pointings in the survey area, with approximately 90\,000 individual beams. However,
the Mopra spectrometer delivers 4096 channels for each of 16 simultaneous spectra
yielding a dataset with nearly 6$\times 10^9$ independant voxels ($l$-$b$-$v$ data
elements). This number is prohibitively large such that searching each individual spectrum
for maser emission by eye becomes impractical. In this paper, we describe the source
finding method used in HOPS to identify \water~masers and compare its performance to
the new source finding program, {\sc Duchamp}.

\section{HOPS data}
Details of how the Mopra observations were undertaken and the resultant data
have been described elsewhere \citep{walsh11}. However, we summarise the relevant
details here for convenience. Data were accumulated in square tiles that are
0.5$^\circ$ on each side. The tiles were oriented in Galactic coordinates and
positioned either at b\,=\,+0.25$^\circ$ or b\,=\,$-$0.25$^\circ$. Each tile was observed
twice, where one observation was scanned in Galactic longitude and the other
in Galactic latitude. This greatly reduces (but does not completely eliminate)
scanning artifacts that appear as strips in the data cubes. Observing each tile
twice also reduces the noise. The resulting data were regridded onto data cubes
with $30\arcsec \times 30\arcsec$ pixels. The Mopra beam at 22.2\,GHz is
approximately 2.2$^\prime$. The final data cubes were typically strips
10$^\circ$ long and 1$^\circ$ wide. The spectral channel width in the data cubes
is 0.45\,\kms. The main \water~maser data cube covers a velocity range of
approximately $-$570 to +1290\,\kms. Each \water~maser cube consists of 
1203 $\times$ 125 $\times$ 2203 pixels and is 1.3\,GB in size.
Since observations were taken under varying weather conditions and telescope
elevations, the noise level varies by a factor close to 2.

\section{Source finding methods}
Here we outline two source finding methods that are generally available
and their applicability to \water~maser finding in HOPS data. We also
describe HOPSfind, which is the method we used to detect \water~masers in HOPS.

\subsection{\sc Clumpfind}
{\sc Clumpfind} \citep{williams94} is an automated source finding method that was designed to
characterise spectral line emission from molecular clouds. Not only does it find
regions of emission, but it also attempts to break up overlapping regions of
emission into individual detections. For observations that show complex, extended
emission, it is very useful in characterising detections by breaking up the emission
along boundaries in an intuitive fashion (eg. \citealt{walsh07}). However, the
program is computationally intensive, with most of the computational power
spent on identifying boundaries of extended emission and separating possibly
overlapping detections. The \water~maser data from HOPS show very little overlap
between individual maser sites. Also, we do not expect any of the maser sites
to be resolved in the Mopra beam. Thus, we regard {\sc clumpfind} as too
inefficient for our purposes.

\subsection{\sc Duchamp}
{\sc Duchamp}\footnote{http://www.atnf.csiro.au/computing/software/duchamp}\footnote{Version 1.1.13}
\citep{whiting11}
is a source finding program designed with large surveys,
particularly those associated with ASKAP, in mind \citep{johnston07}.
It does not require emission
to conform to a particular shape (eg. Gaussian) or size, other than the user-defined
inputs. It works by searching through data cubes for emission above a certain
cutoff level and then merging detections, based on user-defined inputs. The user-defined
inputs that are relevant to this work are the region within the cube to search,
a threshold (expressed in terms of $\sigma$) over which voxels will be counted
towards a detection, the minimum number of pixels to constitute a detection, the minimum
number of channels which constitute a detection and how far away voxels are
allowed to be from each other in order to be considered in a merged detection.

{\sc Duchamp} also offers the option to reconstruct the data cube using the {\em \`{a} trous}
procedure. This procedure allows the reliable detection of fainter sources and the
removal of spurious detections in a noisy data cube by effectively removing some
of the random noise. More details of this method can be found on the {\sc Duchamp}
website\footnotemark[1].

\subsection{HOPSfind}
\begin{figure}[h]
\begin{center}
\begin{tabular}{c}
\includegraphics[scale=0.38, angle=0]{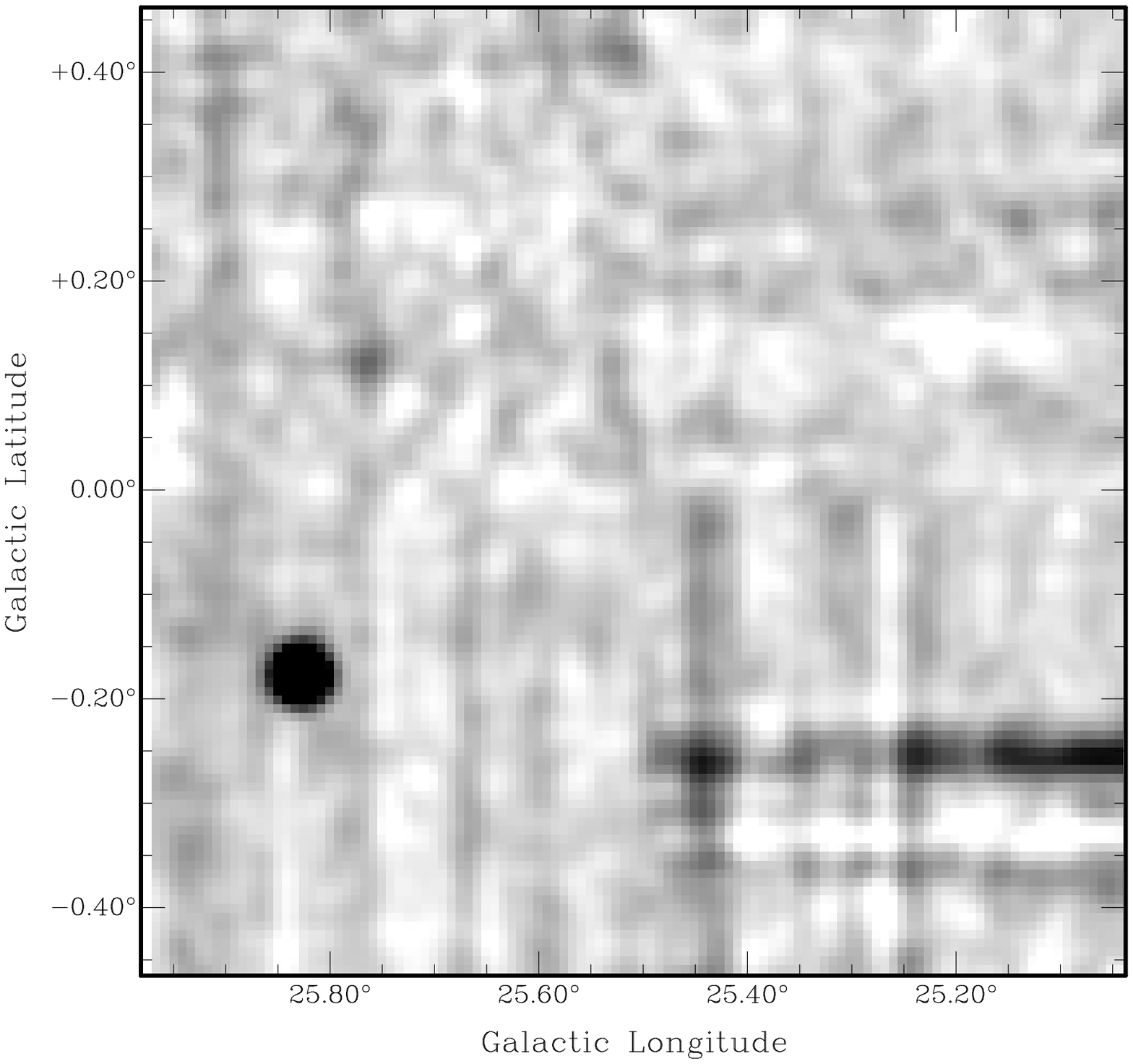}\\
\includegraphics[scale=0.38, angle=0]{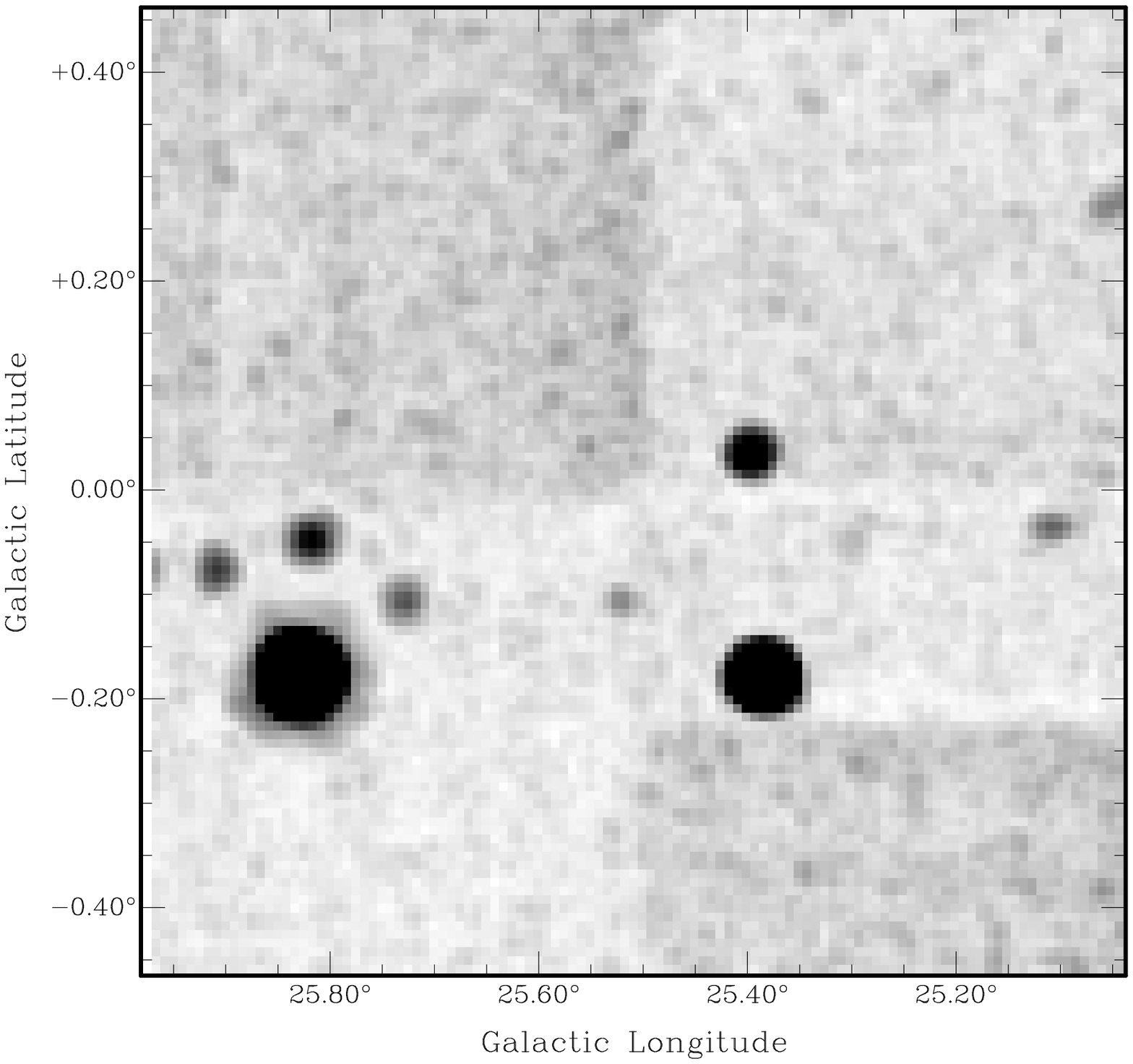}\\
\end{tabular}
\caption{Examples of moment maps used to identify \water~masers in HOPS.
{\bf (Top)}: Integrated intensity, or 0th moment map, where only one maser
can be clearly seen at l\,=\,25.83, b\,=\,$-$0.18. {\bf (Bottom)}: Peak temperature,
or $-$2 moment map, showing 10 masers.}\label{moments}
\end{center}
\end{figure}

The limitation of any source finding method is the amount of time available to
complete the task, so the choice of source finding method is one of efficiency:
how to identify the largest number of real sources without identifying too many
spurious sources in a reasonable time-frame.
During the HOPS pilot observations and data processing \citep{walsh08} it was
realised that each square degree of the Galactic plane is expected to have no more
than about 10 \water~masers, meaning that overlapping masers would be rare. Also,
because the masers are typically spatially unresolved, bright (many Janskys)
and narrow-lined (less than 5\,\kms), it
is relatively easy to manually identify them, compared to weaker, extended emission.
Therefore, we used the following method (which we call ``HOPSfind'') to identify
\water~masers in HOPS:

Each processed data cube was smoothed with a two-dimensional Gaussian kernel of
size $90\arcsec \times 90\arcsec$.
From each data cube, a peak temperature map (PTM) was made. The PTM is made, using
the {\sc miriad} task {\sc moment} and choosing the ``mom=-2'' option. This option
scans through the spectrum at each spatial pixel and finds the brightest emission,
which is then copied to the PTM at the same position. Thus, the PTM is similar
in appearance to
the more commonly used ``0th moment'', or integrated intensity map (IIM). However,
the PTM has one significant advantage over an IIM. The HOPS IIM tends to be dominated
by small errors in the baseline, if it is not completely flat. This is because we have
a broad band (137.5\,MHz across 4096 channels), where the small baseline errors
accumulate in the IIM.
Until recently, radiot elescope observations have been made with narrow bandwidths
and with few channels, where small errors do not dominate in an IIM. It is
with the advent of much broader bandwidths and many thousands of channels that the
IIM becomes less reliable. We also note that attempts to remove the baseline are
difficult because the baseline is usually complicated over the full extent of the band.
A low-order polynomial may be removed from the baseline, improving the result, but small-scale
deviations still tend to be significant in an IIM.
Because this problem is likely to affect many spectral line surveys in
the future, we suggest that PTMs are used in favour of IIMs for locating emission.
Figure \ref{moments} shows a comparison of an IIM (top) and PTM (bottom). In the
IIM, it is possible to clearly identify one maser, but many more are seen in the PTM.

We found that the PTMs are very useful in identifying most of the masers in the
survey. However, the \water~maser spots appear as peaks that are typically broader
than the spectral channel width (0.45\,\kms), so our method can be improved by
averaging adjacent channels together before producing the PTM. Since we found that
the average maser line width is 1.4\,\kms, we average 3$\times$ 0.45\,\kms~channels together.

Maser candidates were then visually checked, using the PTMs and by inspection of
the spectrum and raw data cube at maser candidate positions. A maser candidate was
deemed confirmed if the peak signal was greater than five times the noise level
($5\sigma$) in at least two adjacent channels or if there were at least three
adjacent channels that were 3$\sigma$ or greater in the raw (ie. unaveraged) data.
We also checked for any signals appearing in a single channel above 8$\sigma$, but
found no new masers that did not already satisfy the above requirements.
Using this method, we identified 381 masers (approximately 71\% of all masers
identified).

\water~masers often occur with multiple peaks in the spectrum. Each of these peaks (called
maser spots) are typically found spatially close to each other, on scales of a few
arcseconds or less. The grouping of maser spots is referred to as a maser site. The
spots within each site are expected to be associated with a single site of star formation.
Multiple instances of maser sites are also often found within the same star forming cloud.
Given the Mopra beam, it is not possible to resolve a single maser site or multiple maser
sites that are within the Mopra beam. Thus, if we see multiple peaks within a spectrum at
a single location, we only assign it to a single detection, even though it is possible
that many single detections are comprised of multiple maser sites.

Any weak maser candidates that were not confirmed in the initial list of 381 masers
were reobserved with Mopra, using two minutes on-source plus two minutes off-source
integration time, resulting in a noise level about $\sqrt{2}$ lower than the initial
observations. Any spectral features that were not common in both observations
were discarded as spurious. Unfortunately, because \water~masers are known to vary,
it is possible that some real masers detected in the first observations were below
the detection limit of the second observations and thus discarded.

In all, we identified 540 confirmed \water~masers, using the two-stage method outlined
above.

\section{Source finding with {\sc Duchamp}}
As described above, {\sc duchamp} is an automated source finding method that we can
utilise to compare HOPSfind for detecting \water~masers. {\sc duchamp} uses a single
detection threshold for the entire cube, rather than taking into account local noise
levels, as is naturally done by
the human eye. Therefore, in order to create the best data cube for {\sc duchamp} to
search for sources, it is necessary to further process the data. We average a section
of the \water~maser data cube that is free of line emission (500-1000\kms) and use
the resultant 2-dimensional image as an rms noise map. The full \water~maser cube is
then divided by the noise map to effectively create a signal-to-noise cube, where the
majority of noise variations have been smoothed out. It is this signal-to-noise cube
that we use {\sc duchamp} to search for source.

Successful operation of
{\sc duchamp} requires fine tuning of the input parameters so that as many real sources
are found, with minimal spurious detections. At some level, there is a trade off between
how deep into the noise one chooses to search for weak masers and how much extra time
is required to eliminate spurious detections. This applies equally to any source finding
method. Our most efficient input parameters that we chose for {\sc duchamp} are given
in Table \ref{duckmaps}. These input parameters effectively mean that there must
be emission over at least one beam in a single spectral channel and that a source must
comprise of emission in at least two adjacent channels. This effectively eliminates the
chance of finding masers that appear in only one spectral channel. However, our experience
with HOPSfind found no such masers, so that these requirements become
an effective method to remove spurious sources that may appear in a single channel.
The requirement that there is emission in at least two adjacent channels is similar to the
3-channel averaging used in HOPSfind. We found that using only two adjacent channels and
not three, as well as a cutoff level set at 3$\sigma$, gives results most closely
matched to HOPSfind, allowing for a direct comparison.

\begin{table}[h]
\begin{center}
\caption{Input parameters for {\sc duchamp}.}\label{duckmaps}
\begin{tabular}{lcl}
\hline
Parameter & Value & Description\\
\hline
minPix & 15 & \begin{minipage}{3.4cm}{Must have at least 15 spatial pixels in one velocity channel}\end{minipage}\vspace{2mm}\\
\hline
minChannels & 2 & \begin{minipage}{3.4cm}{Must include emission in at least two channels}\end{minipage}\vspace{2mm}\\
\hline
minVoxels & 26 & \begin{minipage}{3.4cm}{Must have at least 26 pixels in any channel in the cube}\end{minipage}\vspace{2mm}\\
\hline
snrCut & 3 & \begin{minipage}{3.4cm}{Pixels must be at least 3$\times$ the rms noise level}\end{minipage}\vspace{2mm}\\
\hline
flagAdjacent & true & \begin{minipage}{3.4cm}{Channels must be adjacent to be merged}\end{minipage}\vspace{2mm}\\
\hline
\end{tabular}
\medskip\\
\end{center}
\end{table}

\section{Comparison of HOPSfind with {\sc Duchamp}}
\label{compare}
An important consideration in the choice of any source finding method is how
much time is required. We found that to search each square degree for \water~masers,
HOPSfind typically took about 10 minutes so that the entire survey region can be
searched in around 1000 minutes. Most of this time is taken in transcribing the
coordinates of detected masers to a source file and checking correctness of the
transcribed details. The time taken for
{\sc Duchamp} to search each 10-degree strip was approximately 4.5 minutes so the entire
100 square degrees of the survey region was searched in approximately 45 minutes.
Thus, the running time for {\sc duchamp} is significantly less than for HOPSfind. However,
this time does not take into account the time required to set up each method, which
is very short for HOPSfind but typically much longer for {\sc duchamp}. This is due to
time taken to alter input parameters in order to find the most efficient balance
when running {\sc duchamp}. Thus, on small datasets, a manual source-finding method such as HOPSfind
is quicker to implement and complete. On much larger datasets, HOPSfind will take much longer
due to the time taken to catalogue each detection.

Overall the {\sc duchamp} source finding method found 620 detections.
Of the 620 detections identified by {\sc duchamp}, 491 of them overlap with confirmed
masers from HOPSfind. The majority of the remainder are considered spurious detections
after visually checking the candidates, as was done for HOPSfind. A large
fraction of the spurious detections using {\sc duchamp} appear to be artifacts
along the scanning directions used in the observations. Inspection of the raw
data cubes in these instances reveals linear features in either Galactic longitude
or latitude and spurious detections are often made where two such scanning
artifacts meet. Such features are easy for the human eye to identify, but difficult
for an automated method. The scanning artifacts are generally of two types: a `ripple'
in the spectral dimension caused by a poorly subtracted baseline, or a region of
elevated noise.

\begin{figure}[h]
\begin{center}
\includegraphics[scale=0.38, angle=0]{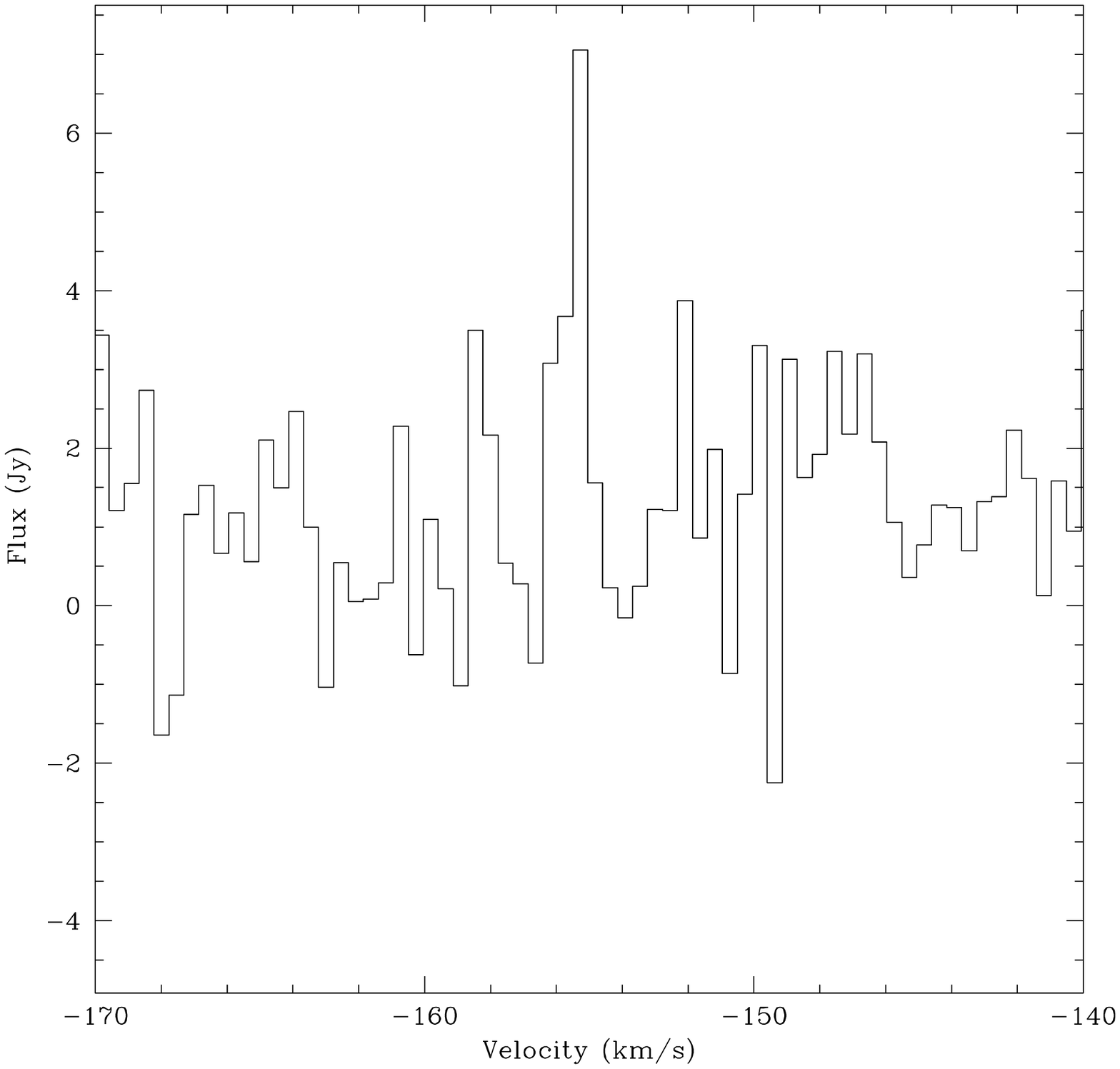}
\caption{Spectrum of the spurious detection made by {\sc duchamp} at
the position (l, b) = (25.43, $-$0.38). The strongest peak is found at
a velocity of $-$155.3\,\kms, with an adjacent channel at $-$155.7\,\kms.}
\label{clump3spec}
\end{center}
\end{figure}

\begin{figure}[h]
\begin{center}
\begin{tabular}{c}
\includegraphics[scale=0.38, angle=0]{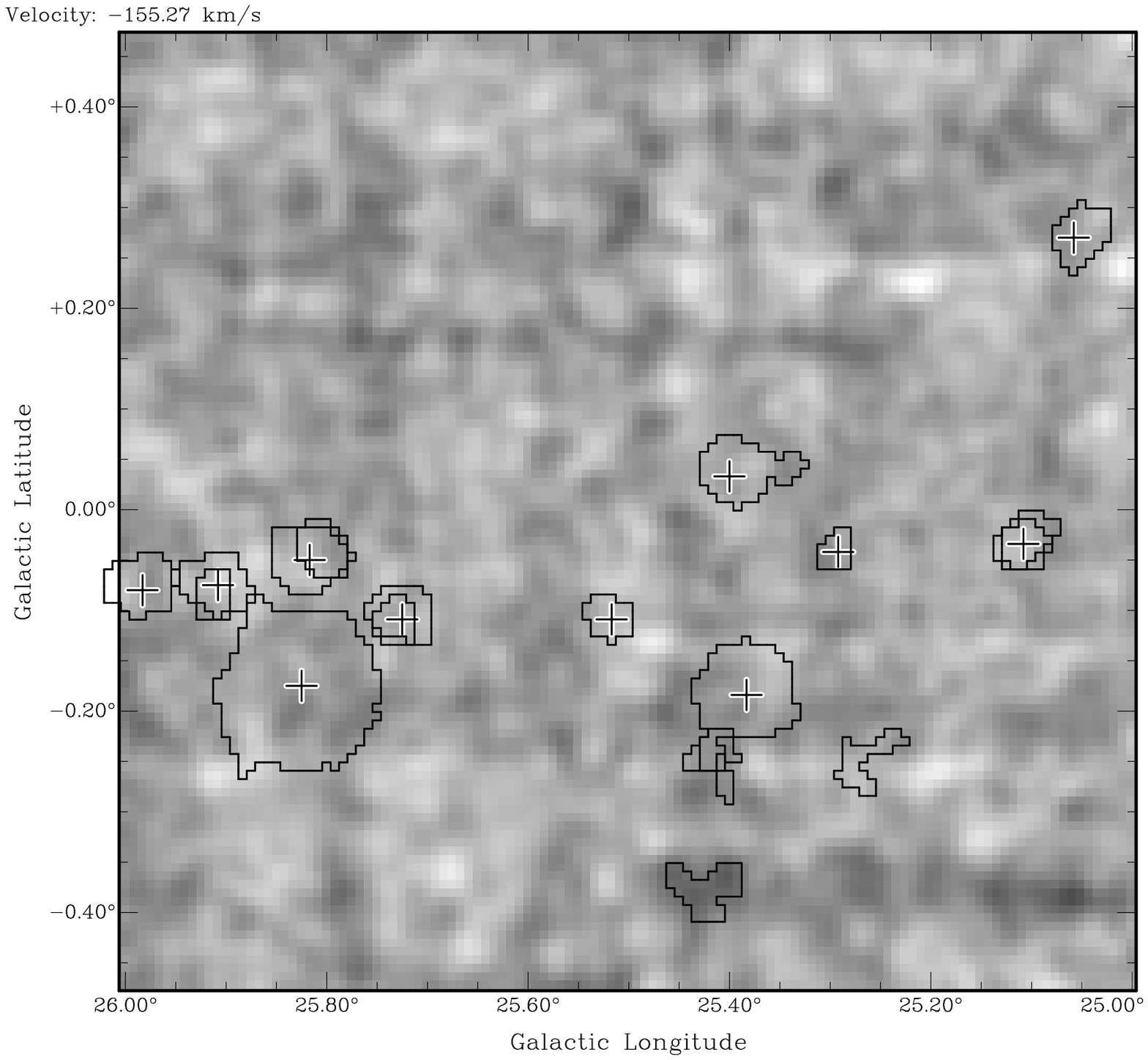}\\
\includegraphics[scale=0.38, angle=0]{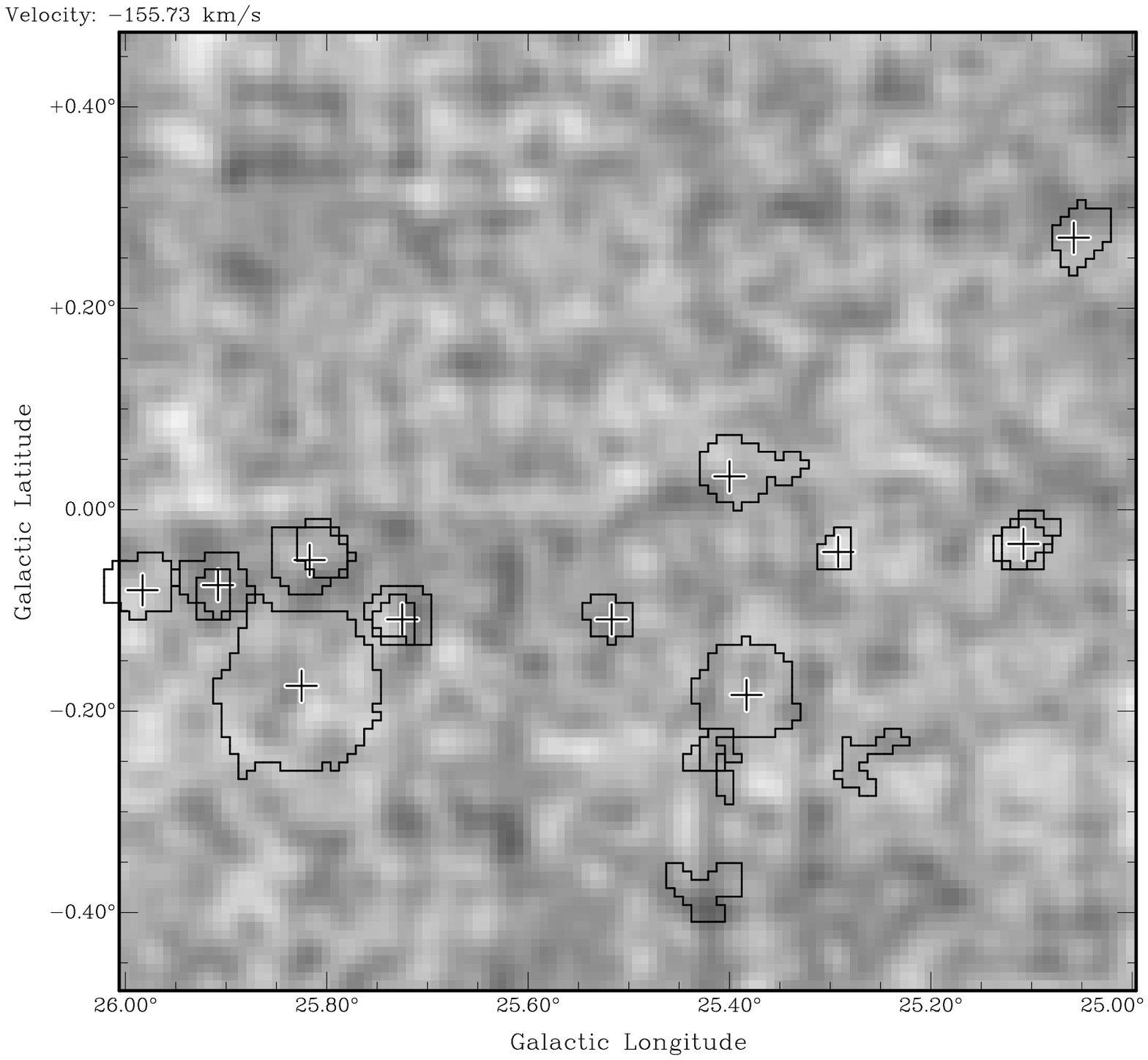}\\
\end{tabular}
\caption{Two adjacent velocity channels are shown in the G025$-$026 region.
The plusses indicate positions of confirmed \water~masers and the regions
outlined with black lines are the results of {\sc duchamp}. Within the
field of view, there are three {\sc duchamp} detections that are not
associated with confirmed \water~masers and are most likely
spurious detections centred on (l,b) = (25.42, $-$0.25), (25.43, $-$0.38)
and (25.28, $-$0.25).}
\label{clump3}
\end{center}
\end{figure}

Examples of spurious detections are shown in Figures \ref{clump3spec} and
\ref{clump3}. Figure \ref{clump3spec} shows a peak in the spectrum at
around $-$155.3\,\kms, consisting of at least two adjacent channels. These
two channels are shown as maps from the data cube in Figure \ref{clump3}.
The positions of confirmed \water~masers are shown in Figure \ref{clump3},
with plus symbols and the location of detections made
by {\sc duchamp} are outlined with black solid lines. As can be seen from the Figure,
most of the {\sc duchamp} detections coincide with confirmed masers. However,
there are three {\sc duchamp} detections which do not coincide with confirmed
\water~masers at (l,b) = (25.42, $-$0.25), (25.43, $-$0.38) and (25.28, $-$0.25).
The {\sc duchamp} detection represented in Figure \ref{clump3spec} is located
at (25.43, $-$0.38). The two channels of the data cube do not show a spatially
confined peak of emission at this position. Instead, both channels show positive
scanning artifacts that go through this position. The channel at
$-$155.3\,\kms~shows a scanning artifact at a Galactic longitude of $-$0.38$^\circ$
and the channel at $-$155.7\,\kms~shows a weaker scanning artifact at a Galactic
latitude of 25.42$^\circ$. Thus, {\sc duchamp} identifies the convergence of these
two artifacts as a detection. Inspection of the data cube at the positions and
velocities of the other two {\sc duchamp} detections that do not coincide with
a confirmed \water~maser indicate that they too are the result of scanning
artifacts.

It may be possible in the future to automatically eliminate some of these
spurious detections by imposing a limit on the maximum size, or number of pixels
that constitute a detection. The maximum size would be set at slightly larger
than the beam FWHM. This is because we assume that \water~maser detections
will always appear unresolved in our observations. Unfortunately, it is not possible
to specify a maximum number of pixels in {\sc duchamp}. Even if this
were possible, care must be used in removing such spurious detections, since very
strong masers will also be identified as spurious, as emission above the detection
threshold will cover a larger area than the beam FWHM. This can be seen in Figure
\ref{clump3} where the {\sc duchamp} detection associated with the brightest
\water~maser, located at (25.83, -0.17), includes pixels that cover an area much
larger than the beam FWHM. It is likely that a two-stage
approach will be most useful: The first stage identifies strong masers above
a high cutoff level, with no limitation on the maximum number of pixels in a
detection. The second stage uses a low cutoff level to search for the weakest
masers and also uses a maximum number of pixels to eliminate spurious detections
much larger than the beam size.

Another approach is to remove scanning artifacts prior to running {\sc duchamp}.
\citet{purcell11} have successfully demonstrated that most baseline ripples may be
removed from HOPS data by fitting low-order polynomials to the line-free channels
so that the performance of {\sc duchamp} is greatly improved.
Scanning artifacts which manifest as regions of elevated noise may be mitigated by
dividing each spectral plane by a `noise-map' to create a signal-to-noise cube,
as described above.

Of the 129 detections found by {\sc duchamp}, but not confirmed as \water~masers, we
find 11 of them that are possibly real detections, based on their size, shape and
spectra, that were not identified using HOPSfind. These
maser candidates await followup observations with Mopra to confirm them as real
masers, or otherwise.

\subsection{Comparison of completeness limits}

An alternate method of comparing our source finding method to {\sc duchamp} is to
test the efficiency of both methods in detecting weak sources that have been
artificially created in a source-free data cube. Note that we used this method in
HOPS to determine the completeness limit of HOPSfind \citep{walsh11}
and details can be found therein. We used the HOPS data within the l=296 to 297
square degree as our template, in which we did not detect any confirmed masers,
but has noise characteristics representative of the full survey. Using the
{\sc miriad} task {\sc imgen}, we injected five sources, with typical \water~maser
characteristics (ie. FWHM of 1.4\,km\,s$^{-1}$), at random positions and
velocities into the data cube and searched for detections using HOPSfind and
{\sc duchamp}. This was repeated twenty times so that in total 100 sources were
searched for. We varied the intensity of the generated sources to four values,
equivalent to a peak of 5.5, 6.7, 8.4 and 11.1\,Jy. The rms noise level in the
data cube was 1.5\,Jy. We also performed the source-finding on a signal-to-noise
cube, as described in \S\ref{compare}.

\begin{figure}[h]
\begin{center}
\includegraphics[scale=0.38, angle=0]{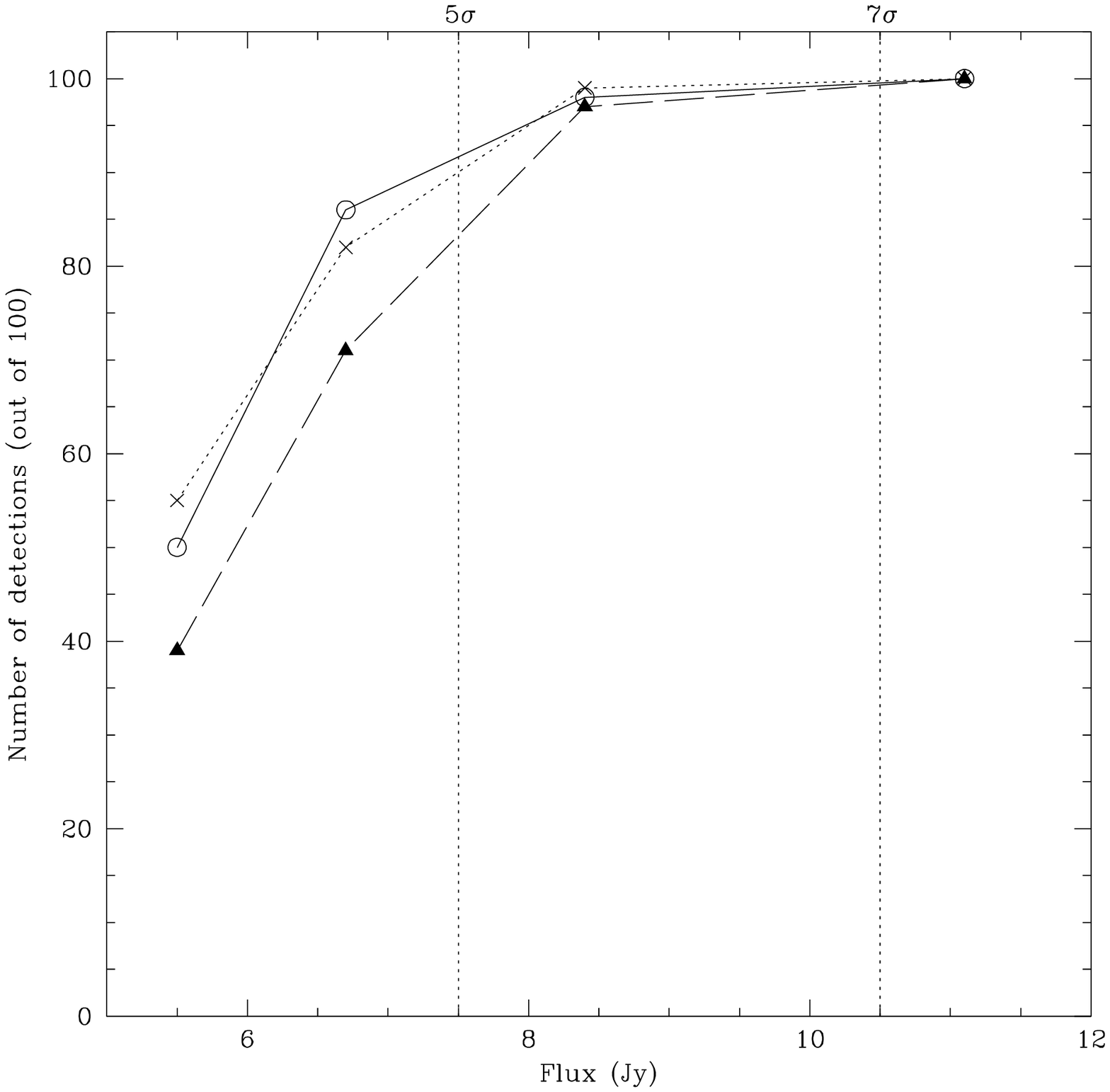}
\caption{The number of successful detections of generated sources, out of a possible
100, is plotted against the peak flux density of the generated source, using 
HOPSfind (solid line and open circles), {\sc duchamp} (dashed
line and filled triangles) and {\sc duchamp} with {\em \`{a} trous} wavelet reconstruction
(dotted line and crosses). The vertical lines represent the 5 and 7 times the rms
noise level. All methods perform similarly, with the success of detection
using {\sc duchamp} on its own slightly lower than the other two.}
\label{complete}
\end{center}
\end{figure}

Figure \ref{complete} shows the success rate of detecting the generated sources
in the data cube for HOPSfind and {\sc duchamp}. We find that the methods
have very similar success rates in detecting the generated sources, with HOPSfind
detecting around 20\% more sources than {\sc duchamp}. We have used the same cutoff
threshold for {\sc duchamp} and in this one square degree data cube, it only detected
one spurious source. It would be possible to lower the cutoff and
increase the detection rate to more closely match that of HOPSfind,
but this introduces more spurious detections that must later be removed.
We note that {\sc duchamp} can also perform an {\em \`{a} trous} wavelet
reconstruction, which determines the amount of structure present and then removes
random noise from the cube. This can significantly improve the source detection statistics
for {\sc duchamp}. {\sc Duchamp}, with the {\em \`{a} trous} wavelet reconstruction
incorporated is shown in Figure \ref{complete} as the dotted line with crosses. For this
implementation, we used three dimensions for reconstruction (ie. the full cube is
reconstructed in one go) and a reconstruction threshold of 2.5. It
can be seen that with the {\em \`{a} trous} wavelet reconstruction method, {\sc duchamp}
does indeed perform better, with results similar to HOPSfind.

We note that \water~masers exhibit a range of line widths, whereas to estimate completeness
limits, we have only injected sources with a single line width. However, we do not expect
this will significantly affect our completeness comparison. This is because firstly 85\% of
all masers have linewidths within a factor of two from the average 1.4\,km\,s$^{-1}$. 
Secondly, we expect both HOPSfind and {\sc duchamp} to perform similarly in detecting 
either unusually small-FWHM or large-FWHM masers. Thus, whilst the absolute completeness
limits may be affected by the linewidth of the maser emission, we do not expect there
to be significant differences between the relative performance of HOPSfind and {\sc duchamp}.

These results generally show that the final results from each source finding method are
very similar. Our intention is not to show that one source finding method is better than any
other, but we imply that the efficiency of any source finding method is naturally limited
by the noise in the data. Thus, well-tuned source finding methods will produce similar
results and the choice of source finding method should be based on the ease of use, as well
as other functionality, such as the ability to characterise extended sources.

\section{Data Visualisation}
Cross-matching HOPS datasets with complementary Galactic plane surveys
is a key step towards achieving the science goals of the project. For
example, matching NH$_3$ clouds and H$_2$O masers to features in the
mid-infrared Midcourse Space Experiment \citep{egan96}
makes it possible to determine the physical properties of star forming regions.
To facilitate visualising the data we used the {\it
  Chromoscope}\footnote{http://www.chromoscope.net
  https://github.com/slowe/Chromoscope/} software to 
directly compare large-scale HOPS images with other surveys.  {\it
  Chromoscope} is an all-sky image-viewer which allows the user to
easily move around the sky, fade between co-registered images and
zoom-in to inspect features of interest. Because {\it Chromoscope} is
javascript based it will run in any modern web-browser and is an ideal
tool for publicising images online to the general public. However, it
does not require an internet connection and has been used to present
data at conferences, and to select interesting objects for follow-up
studies. The software supports Keyhole Markup Language (KML) files,
used to annotate positions of objects in the popular {\it Google
  sky}\footnote{http://www.google.com/sky}. Tools are also provided to
convert astronomical catalogues to KML format for overplotting on the
images. If an internet connection is available the {\it Chromoscope}
can utilise the astronomical name resolver `{\scriptsize
  lookUP}'\footnote{http://www.strudel.org.uk/lookUP/} to find and
display specific astronomical objects. The HOPS data is presented on
{\it Chromoscope} at http://www.ast.leeds.ac.uk/hops/chromoscope.

\section{Conclusions}

We have compared two methods for finding \water~maser sources in the HOPS data.
First, we used HOPSfind, which is a manual method of searching for signals
in a peak temperature map, that is binned in such a way as to give the greatest
contrast to typical \water~maser profiles. Second, we used {\sc duchamp}, which is
an automated method of finding masers.

Overall, we find the two methods of detecting sources to perform similarly well.
In order to detect weak masers, it is always necessary to also accumulate spurious
detections, which must later be assessed through followup observations. We find some minor
limitations to the {\sc duchamp} method because spurious detections that are the result
of scanning artifacts are not flagged out. These limitations could be overcome in the
future by rejecting candidate detections that do not conform to the beam size and shape
or by running {\sc duchamp} on a signal-to-noise cube which has had lower-order polynomials
subtracted from the baselines.

The {\sc duchamp} method finds proportionately more spurious sources than HOPSfind, given
the same detection limit. However, when including the {\em \`{a} trous} wavelet reconstruction
method, we find similar detection efficiencies. This is because the detection efficiency
is not strongly dependant on the method used, but rather the noise characteristics of
the data. Thus, the choice of source finding method should be based on ease of use and
other features, rather than detection efficiency.
We conclude that HOPSfind is more appropriate to detecting \water~masers in HOPS, since
they are few and far between, as well as being spatially unresolved, so that no extended
emission needs to be characterised. For surveys of extended emission, such as the ammonia
inversion transitions detected in HOPS \citep{purcell11}, {\sc duchamp} is a better source
finding method to use, since it automatically characterises the extended emission. It is
expected that in future surveys, such as those planned with ASKAP, many more sources will
be detected, making a manual method like HOPSfind impractical. For example, GASKAP expects
to find around 15\,000 OH masers. In addition to this, extended
emission from thermal line and continuum emission needs to be characterised, making
{\sc duchamp} the more appropriate source finder.

\section*{Acknowledgments} 
The Mopra radio telescope is part of the Australia Telescope National Facility which
is funded by the Commonwealth of Australia for operation as a National Facility managed
by the CSIRO. The authors would like to thank Stuart Lowe from Las Cumbres Observatory Global
Telescope Network who wrote {\it Chromoscope} and helped greatly in the creation of
the HOPS {\it Chromoscope} website.



\begin{thebibliography}{}
\bibitem[Barlow et al.(1996)Barlow et al.]{barlow96}
Barlow M.~J. et al. 1996, A\&AL, 315, 341
\bibitem[Caswell et al.(2010)Caswell et al.]{caswell10}
Caswell, J.~L. et al. 2010, MNRAS, 404, 1029
\bibitem[Cheung, Rank \& Townes(1969)Cheung et al.]{cheung69}
Cheung A.~C., Rank D.~M., Townes C.~H. 1969, Nature, 221, 626
\bibitem[Claussen et al.(1996)Claussen et al.]{claussen96}
Claussen M.~J., Wilking B.~A., Benson P.~J., Wootten A., Myers P.~C.,
Terebey, S. 1996, ApJS, 106, 111
\bibitem[Claussen et al.(1984)Claussen et al.]{claussen84}
Claussen M.J. et al., 1984, ApJL, 285, 79
\bibitem[Dickinson(1976)Dickinson]{dickinson76}
Dickinson D.~F. 1976, ApJS, 30, 259
\bibitem[Egan \& Price(1996)]{egan96}
Egan, M.~P. \& Price, S.~D. 1996, AJ, 112, 2862
\bibitem[Forster \& Caswell(2000)]{forster00}
Forster J.~R., Caswell J.~L., 2000, ApJ, 530, 371
\bibitem[Forster \& Caswell(1999)Forster \& Caswell]{forster99}
Forster J.~R., Caswell J.~L. 1999, A\&AS, 137, 43
\bibitem[Gundermann(1965)Gundermann]{gundermann65}Gundermann E. 1965, PhD Thesis,
Harvard Univ., Cambridge, Mass., USA
\bibitem[Hinkle \& Barnes(1979)Hinkle \& Barnes]{hinkle79}
Hinkle K.~H., Barnes T.~G. 1979, ApJ, 227, 923
\bibitem[Johnston et al.(2007)Johnston et al.]{johnston07} 
Johnston, S. et al. 2007, PASA, 24, 174
\bibitem[Miranda et al.(2001)Miranda et al.]{miranda01} 
Miranda L.~F., G\'{o}mez Y., Anglada G., Torrelles, J.~M., 2001,
Nature, 414, 284
\bibitem[Purcell et al.(2011)Purcell et al.]{purcell11}
Purcell, C.~R. et al. 2011, MNRAS {\em submitted}
\bibitem[Voronkov et al.(2005)Voronkov et al.]{voronkov05}
Voronkov, M.~A., Sobolev, A.~M., Ellingsen, S.~P., Ostrovskii, A.~B. 2005, MNRAS, 362, 995
\bibitem[Walsh et al.(2011)Walsh et al.]{walsh11}
Walsh, A.~J. et al. 2011, MNRAS, 416, 1764
\bibitem[Walsh et al.(2008)Walsh et al.]{walsh08}
Walsh, A.~J., Lo N. Burton M.~G., White G.~L., Purcell C.~R.,
Longmore S.~N., Phillips C.~J., Brooks K.~J. 2008, PASA, 25, 105
\bibitem[Walsh et al.(2007)Walsh et al.]{walsh07}
Walsh, A.~J., Myers, P.~C., Di Francesco, J., Mohanty S., Bourke T.~L., Gutermuth R., Wilner D.
2007, ApJ, 655, 958
\bibitem[Walsh et al.(1998)]{walsh98} Walsh A.J.,
Burton M.G., Hyland A.R., Robinson G., 1998, MNRAS, 301, 640
\bibitem[Weaver et al.(1965)Weaver et al.]{weaver65}
Weaver H., Williams D.~R.~W., Dieter N.~H., Lum W.~T. 1965, Nature, 208, 29
\bibitem[Whiting(2011)Whiting]{whiting11} Whiting, M. 2011, MNRAS, {\em submitted}
\bibitem[Williams, de Geus \& Blitz(1994)Williams et al.]{williams94}
Williams, J.~P., de Geus E.~J., Blitz L. 1994, ApJ, 428, 693

\end{thebibliography}
\end{document}